\documentclass[aps,prl,twocolumn,showpacs,superscriptaddress]{revtex4-2}
\usepackage{graphicx}
\usepackage{amsmath}
\usepackage{amssymb}
\usepackage{colordvi}
\usepackage{mathrsfs}
\usepackage{bm}
\usepackage{verbatim}
\usepackage{dcolumn}
\usepackage{bm}
\usepackage{epsfig}
\usepackage{subfigure}
\usepackage{multirow}
\usepackage[colorlinks=true,allcolors=blue]{hyperref}
\usepackage{pifont}

\begin{document}

\title{Quantum Anomalous Hall Effect in Flat Bands with Paramagnetism}
\author{Yedi Shen$^\dagger$}
\affiliation{International Center for Quantum Design of Functional Materials, and Department of Physics, University of Science and Technology of China, Hefei, Anhui 230026, China}
\affiliation{Hefei National Laboratory, University of Science and Technology of China, Hefei 230088, China}
\author{Sanyi You$^\dagger$}
\affiliation{International Center for Quantum Design of Functional Materials, and Department of Physics, University of Science and Technology of China, Hefei, Anhui 230026, China}
\affiliation{Hefei National Laboratory, University of Science and Technology of China, Hefei 230088, China}
\author{Zhenhua Qiao}
\email[Correspondence author:~~]{qiao@ustc.edu.cn}
\affiliation{International Center for Quantum Design of Functional Materials, and Department of Physics, University of Science and Technology of China, Hefei, Anhui 230026, China}
\affiliation{Hefei National Laboratory, University of Science and Technology of China, Hefei 230088, China}
\author{Qian Niu}
\affiliation{International Center for Quantum Design of Functional Materials, and Department of Physics, University of Science and Technology of China, Hefei, Anhui 230026, China}
\date{\today}

\begin{abstract}
Quantum anomalous Hall effect has been widely explored in both ferromagnetic and antiferromagnetic systems. Here, we propose an interaction-driven paramagnetic quantum anomalous Hall effect emerging in the Fermion-Hubbard model on a dice lattice with weak spin-orbit coupling. Based on exact diagonalization calculations, the time-reversal symmetry breaking in the ground state is evidenced by nonuniform loop currents between nearest-neighbor sites. The many-body ground state possesses a Chern number of $\mathcal{C}=2$ or $6$, and strong correlation effects in the half-filled flat bands lead to a well-defined first excitation gap and a clear insulating gap, ensuring the robustness against thermal fluctuations and external perturbations. The interplay between spin-orbit coupling and Hubbard interaction allows tunability of various magnetic ground states, generating a rich phase diagram with competing ferromagnetic, antiferromagnetic, and paramagnetic orders.
\end{abstract}

\maketitle

\textit{Introduction}.---Quantum Hall effect is the first topological phase observed in a two-dimensional electronic system subjected to low temperature and strong magnetic field, characterized by the quantization of 
Hall conductance~\cite{PhysRevLett.45.494,PhysRevB.23.5632,PhysRevLett.49.405}. In the honeycomb lattice, quantized Hall conductance can also arise from a staggered magnetic flux that breaks time-reversal symmetry while preserving zero net flux, i.e., without an external magnetic field---a hallmark signature of the 
quantum anomalous Hall effect (QAHE)~\cite{PhysRevLett.61.2015}. Ferromagnetism with intrinsic time-reversal symmetry breaking provides an ideal platform to realize the QAHE by magnetic proximity 
effects~\cite{Science329,PhysRevB.82.161414,PhysRevB.87.085431,CoshareScience.02.01}, magnetic modulation doping 
techniques~\cite{PhysRevLett.101.146802,science.340.167,FOP.13.137308}, and 
electronic correlation effects~\cite{science.365.605, science.367.900, 
Nature.588.610,PhysRevLett.126.117602,PhysRevB.108.L161112,Commun.Phys.6.240,PhysRevLett.129.036801}. Antiferromagnetic QAHE, characterized by zero net spin magnetization, has also been proposed in recent 
years~\cite{Nature.576.416,PhysRevLett.122.107202,npjQuantum 
Mater.5.54,PhysRevB.100.121103,SciAdv.5.eaaw5685,PhysRevX.9.041038,PhysRevLett.122.206401,PhysRevLett.123.096401,PhysRevB.101.201408,NanoLett.20.2609,PhysRevB.101.161109,PhysRevB.101.161113,PhysRevB.102.241406,PhysRevLett.124.126402,PhysRevB.103.245403,PhysRevLett.134.116603}.
The diminishing dependence of QAHE on magnetic field and spin magnetism—from initial magnetic-field control to zero net flux, and further from spin-polarized ferromagnets to fully compensated antiferromagnets—naturally motivates a fundamental question: does a paramagnetic phase support QAHE? Despite several theoretical proposals of interaction-driven QAHE 
have been presented in the spinless extended Fermion-Hubbard 
model~\cite{PhysRevLett.100.156401,PhysRevLett.117.096402,PhysRevB.98.125144,PhysRevB.98.205146,PhysRevB.106.205105}, the realization of QAHE in spinful correlated paramagnetic systems has yet to be considered.
Enabled by the elimination of spin magnetism, the paramagnetic QAHE stands out as an optimal candidate with robust high-temperature performance and remarkable resilience against external perturbations.

\begin{figure*}[t]
  \centering 
  \includegraphics[width=18cm]{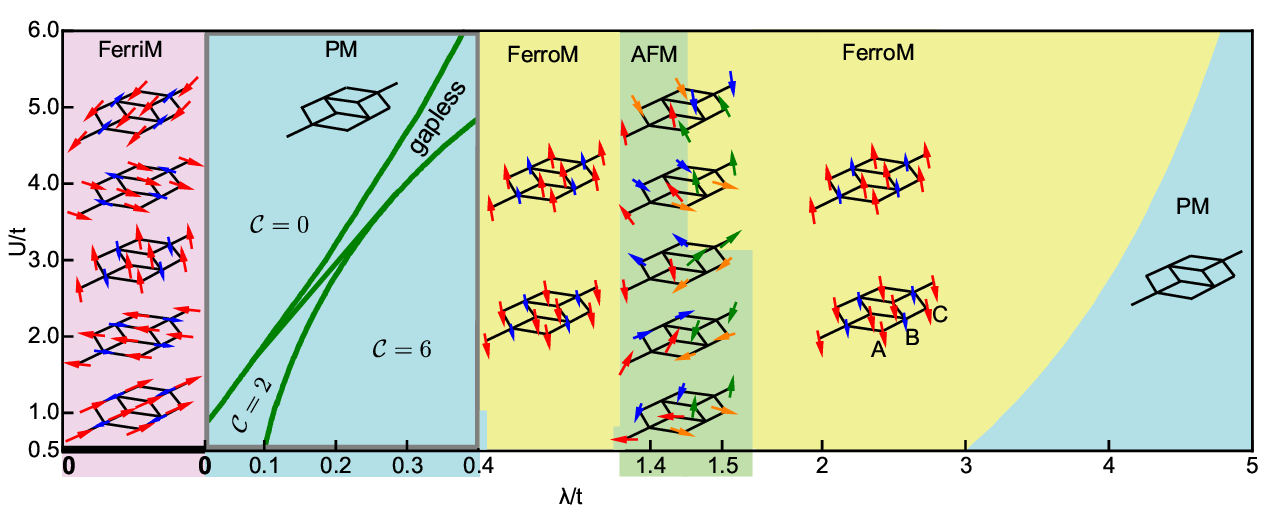}
  \caption{Phase diagram of ground states in the Fermion-Hubbard model on a 
  dice lattice with SOC, shown in the ($\lambda$, $U$) plane. By changing the 
  strengths of SOC at a fixed Hubbard interaction, the initial ferrimagnetic 
  (FerriM) phase transits into paramagnetic (PM), ferromagnetic (FerroM), 
  antiferromagnetic (AFM), ferromagnetic, and paramagnetic phase, respectively. 
  The paramagnetic phase with weak SOC supports Chern 
  number $\mathcal{C}=2$ or $6$. Inset: Magnetic configurations of the 
  degenerate ground states are obtained from exact diagonalization 
  calculations. Under our configurations, the degeneracy of ground 
  states for ferri/para/ferro/antiferro/ferro/paramagnetic phase is 
  5/1/2/5/2/1, respectively. $A$, $B$, and $C$ label the three 
  basis sites in the unit cell.}
  \label{fig1}
\end{figure*}

In this Letter, we propose the paramagnetic QAHE in a Fermion-Hubbard model based on a dice lattice with weak spin-orbit coupling (SOC) in the half-filled flat bands. Our exact diagonalization calculations performed on a 2×2 unit-cell cluster uncover a time-reversal-symmetry breaking ground state, supporting nonuniform loop currents between nearest-neighbor sites. The topological nature of the many-body ground state is characterized by a nonzero Chern number of $\mathcal{C}=2$ or $6$. Considering the strong correlation effect in the half-filled flat bands, we demonstrate a well-defined first excitation gap between the ground state and the first excited state, combined with a clear Mott insulating gap around the Fermi level. Several ground state properties, such as energy, degeneracy, double occupancy, single-particle excitation spectrum, spin-dependent density, spin magnetic moment, and spin-spin correlation are systematically investigated. By tuning the strengths of SOC at certain fixed Hubbard interaction, the initial ferrimagnetic phase gradually transits into the paramagnetic, ferromagnetic, antiferromagnetic, ferromagnetic, and paramagnetic phase, respectively [see Fig.~\ref{fig1}].

\textit{Model and Methods.}---The tight-binding Hamiltonian of Fermion-Hubbard model based on a dice lattice with SOC can be written as~\cite{PhysRevB.84.241103}:
\begin{equation}
	\begin{aligned}
		H_{0} &=-\sum_{\langle i j\rangle\alpha}\left(t c_{i \alpha}^{\dagger} c_{j \alpha}+\text { H.c. }\right)-\sum_{i \in \mathcal{V}_B} \epsilon n_i +\sum_i U n_{i \uparrow} n_{i \downarrow}\\
		&\pm\sum_{\langle i j\rangle\alpha \beta}\left[\mathrm{i} \lambda  \hat{\mathbf{e}}_{z}\cdot (\boldsymbol{\sigma}_{\alpha\beta}\times\mathbf{d}_{ij})c_{i \alpha}^{\dagger} c_{j \beta}+\text { H.c. }\right],
	\end{aligned}
\end{equation}
where $t$ is the nearest-neighbor hopping parameter, $\epsilon$ is the mass term on the sublattice $B$, $\lambda$ is the strength of SOC with $\mathbf{d}_{ij}$ represents a unit vector pointing from site $j$ to site $i$, and $\boldsymbol{\sigma}$ are the Pauli matrices for spin degree of freedom. $c_{i\alpha}^{\dagger}(c_{i\alpha})$ is the creation (annihilation) operator with spin $\alpha$($\uparrow$ or $\downarrow$) at site $i$. The electric field induced by charge transfer between sublattice $B$ and sublattice $A(C)$ is oriented along the $\hat{\mathbf{e}}_{z}(-\hat{\mathbf{e}}_{z})$ direction. $n_{i}=\sum_{\alpha}c_{i\alpha}^{\dagger}c_{i\alpha}$ is the electron density at site $i$, and $U$ is the strength of onsite density-density repulsion interaction. Hereafter, we set $t=1$ and $\epsilon/t=0.6$ for simplicity. To construct a comprehensive magnetic phase transition diagram, we study a broad range of SOC and Hubbard interaction strengths, i.e., $\lambda/t \in \left[0,5\right]$ and $U/t\in \left[0.5,6\right]$~\cite{NC13.4915,PhysRevLett.123.157601}. 

To investigate the characteristics of the degenerate ground states induced by SOC and Hubbard interaction in the half-filled flat bands, we perform exact diagonalization 
calculations~\cite{ComputerPhysicsCommunications2017,ComputerPhysicsCommunications2024}. While the total spin magnetization along $z$-direction is not a conserved physical quantity, we study a $2 \times 2$ unit-cell cluster of the dice lattice (12 sites) under periodic boundary conditions, where the Hilbert space dimension reaches approximately $10^6$. The finite cluster can capture essential long-range magnetic orders, such as 
ferromagnetism and antiferromagnetism~\cite{PhysRevB.37.7359, 
PhysRevB.46.11779, Eur.Phys.J.D.52.159, PhysRevX.5.041041}.

\begin{figure*}[t]
  \centering 
  \includegraphics[width=18cm]{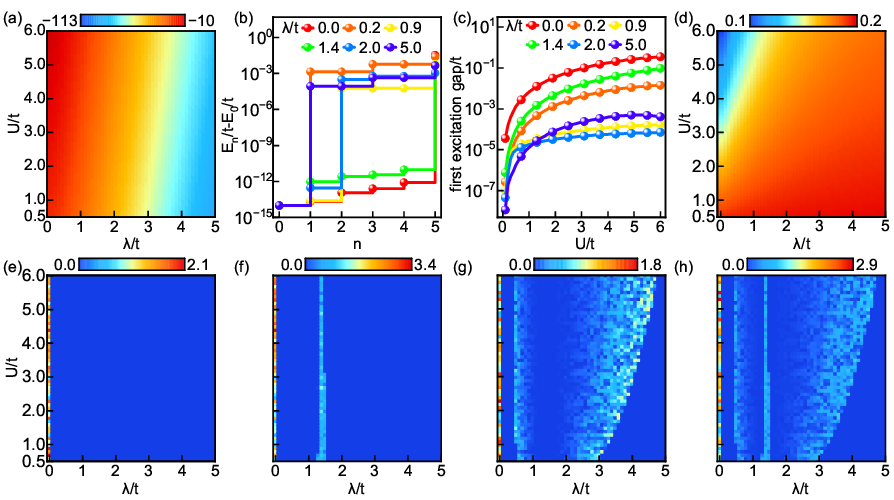}
  \caption{(a) Energy of the ground state. (b) At a fixed Hubbard interaction strength, $U/t=2$, energies of six low-energy excited states are measured from the ground state energy with different strengths of SOC. (c) The first excitation gap varies with the Hubbard interaction strength. (d) Double occupancy possibility of the ground state. Total spin magnetic moment in the $x-y$ plane $\sqrt{|\sum_{i=1}^{12}\langle S_i^x\rangle |^2+ |\sum_{i=1}^{12}\langle S_i^y\rangle|^2}$ (e), $\sqrt{(\sum_{i=1}^{12}|\langle S_i^x\rangle|)^2+ (\sum_{i=1}^{12}|\langle S_i^y\rangle|)^2}$ (f). Total spin magnetic moment in $z$-direction $|\sum_{i=1}^{12}\langle S_i^z\rangle|$ (g), and $\sum_{i=1}^{12}|\langle S_i^z\rangle|$ (h). The unit of spin magnetic moment is $\hbar$.}
  \label{fig2}
\end{figure*}

\textit{Properties of Ground States.}---By using the exact diagonalization method, one can calculate the energy of ground states [see Fig.~\ref{fig2}(a)]. At a fixed Hubbard interaction strength, the increased SOC strength drives bandwidth broadening, consequently lowering the ground state energy. At a fixed SOC strength, the increased Hubbard interaction strength elevates the Fermi level, resulting in a higher ground state energy. At a strong Hubbard interaction of $U/t=2$, in our $2\times2$ unit-cell cluster of the dice lattice, the ground state 
degeneracy changes to 5, 1, 2, 5, 2, and 1 with increasing SOC strength, corresponding to ferri-, para-, ferro-, antiferro-, ferro-, and para-magnetic phases, respectively [see Fig.~\ref{fig2}(b)]. Given the strong electronic correlations in the half-filled flat bands, our exact diagonalization calculations indicate a well-defined first excitation gap between the ground state and the excited states [see Fig.~\ref{fig2}(c)]. Although the onsite repulsive Hubbard interaction prevents double occupancy of electrons with opposite spins at a single site, the SOC counteracts this effect by spin-flip processes. To quantify the possibility of opposite-spin double occupancy, we calculate the expectation value of the double occupancy operator, which is defined as 
$\frac{1}{12}\sum_{i=1}^{12}\langle n_{i\uparrow}n_{i\downarrow}\rangle$. Our 
results show that the enhanced SOC elevates the double occupancy, while stronger Hubbard interaction suppresses it [see Fig.~\ref{fig2}(d)]. Electrons tend to accumulate at $B$ sublattices due to a smaller mass term, i.e., $\langle 
n_{B\alpha} \rangle > \langle n_{A\alpha}\rangle = \langle n_{C\alpha}\rangle$, 
while both SOC and Hubbard interaction can counteract this effect [see Supplemental Material~\cite{SupplementalMaterial} for more details].

To confirm the magnetic configuration of the ground states, we calculate the 
total spin magnetic moments in the $x-y$ plane and $z$-direction. According 
to the general Lieb’s theorem~\cite{PhysRevLett.62.1201, PhysRevLett.72.1280}, 
in the absence of SOC, the Fermion-Hubbard model on the half-filled flat bands 
exhibits a ferrimagnetic ground state, and the degeneracy of 
ground states is 5 in our $2 \times 2$ unit-cell cluster. At weak 
Hubbard interaction strengths ($U/t\le 0.4$), the spin magnetic moments 
tend to be in the $x-y$ plane. In this region, the differences among the lowest 
energies are so small that we cannot distinguish the ground energy from the 
first excited state. With the increase of Hubbard interaction strength ($U/t>0.4$), the spin magnetic moments tend to align along $z$-direction. At a stronger Hubbard interaction, by increasing the SOC strength, the initial ferrimagnetism gradually transits into para-, 
antiferro-, and para-magnetism in the $x-y$ plane, respectively [see 
Fig.~\ref{fig2}(e)(f)]. In $z$-direction, the initial ferrimagnetism 
gradually transits into para-, ferro-, para-, antiferro-, para-, 
ferro-, and para-magnetism, respectively [see Fig.~\ref{fig2}(g)(h)]. By 
combining the spin magnetic moments in $x-y$ plane and 
$z$-direction, we show that the initial ferrimagnetic 
phase transits into para-, ferro-~\cite{PhysRevB.84.241103,PhysRevB.102.045105}, 
antiferro-, ferro-, and para-magnetic phases, respectively. In 
both ferromagnetic and antiferromagnetic phases, the spin magnetic moments 
keep equivalent magnitudes in both sublattices $A$ and $C$, notably more significant 
than those in sublattice $B$. All degenerate ferromagnetic ground states are 
collinear, while all degenerate antiferromagnetic ground states are 
noncollinear. The emergence of the antiferromagnetic phase corresponds to the 
spontaneous translational symmetry breaking.

\begin{figure}[t]
  \includegraphics[width=1.0\linewidth]{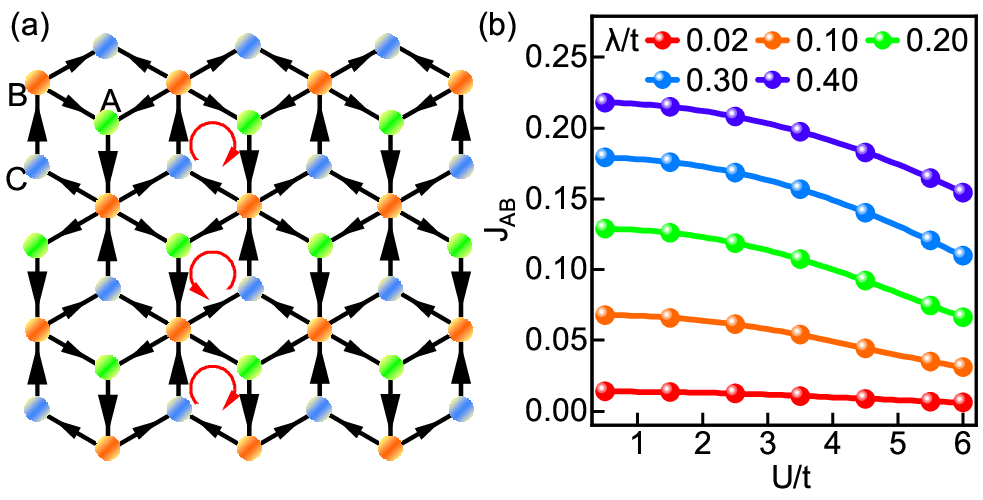}
  \caption{(a) Nonuniform currents in real space show the spontaneous 
  time-reversal symmetry breaking. Black arrows indicate the current 
  directions, whose lengths map the magnitudes of currents on edges. 
  Green/Orange/Blue disks represent sublattice $A$/$B$/$C$. Two kinds of 
  staggered magnetic fluxes exist in the ground state of the paramagnetic phase 
  while preserving zero net flux per unit cell. (b) The magnitudes of currents 
  change with the strength of Hubbard interaction for different SOC.}
  \label{fig3}
\end{figure}

\textit{Spontaneous Time-Reversal Symmetry Breaking.}---From above results, the degeneracy of the paramagnetic ground state is 1. To investigate the potential time-reversal symmetry breaking of the ground state, we calculate the emergent local currents, which are defined as~\cite{PhysRevLett.117.096402,PhysRevB.98.125144}:
\begin{equation}
\begin{aligned}
J_{i j}&=i\left(\langle c_{i \uparrow}^{\dagger} c_{j \uparrow}\rangle-\langle c_{j \uparrow}^{\dagger} c_{i \uparrow}\rangle+\langle c_{i \uparrow}^{\dagger} c_{j \downarrow}\rangle-\langle c_{j \downarrow}^{\dagger} c_{i \uparrow}\rangle\right.\\
& \left.+\langle c_{i \downarrow}^{\dagger} c_{j \uparrow}\rangle-\langle c_{j \uparrow}^{\dagger} c_{i \downarrow}\rangle+\langle c_{i \downarrow}^{\dagger} c_{j \downarrow}\rangle-\langle c_{j \downarrow}^{\dagger} c_{i \downarrow}\rangle\right),
\end{aligned}
\label{equation3}
\end{equation}
where $i$ and $j$ denote the sites connected by the nearest-neighbor bonds. As illustrated in Fig.~\ref{fig3}(a), the local currents exhibit a nonuniform distribution in the real space, signaling a spontaneous breaking of time-reversal symmetry. Most importantly, local currents form a loop structure in the 
counterclockwise and clockwise directions. The staggered magnetic flux in each 
unit cell averages out to be zero as similarly expected from the Haldane model. At a fixed SOC 
strength, the current magnitudes decrease with the increase of Hubbard interaction 
strength. Conversely, at a fixed Hubbard interaction strength, the magnitudes 
of currents demonstrate an increasing trend with the increase of SOC strength [see 
Fig.~\ref{fig3}(b)]. It is noteworthy that the emergent currents are only 
generated by opposite-spin electron contributions, which means $\langle 
c_{i\alpha}^{\dagger}c_{j\alpha}\rangle-\langle 
c_{j\alpha}^{\dagger}c_{i\alpha}\rangle = 0$ in Eq.~(\ref{equation3}). 

\begin{figure}[t]
  \includegraphics[width=1.0\linewidth]{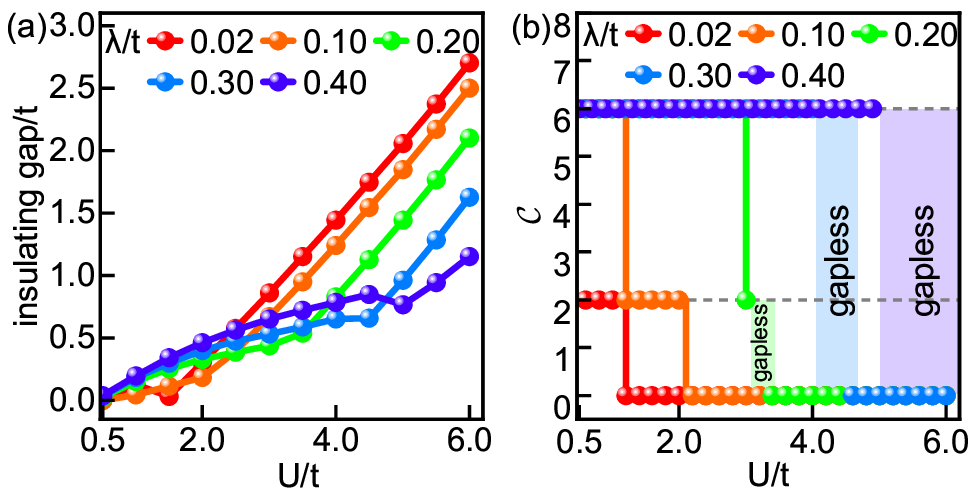}
  \caption{(a) The insulating gap of Mott insulator varies with the strength of 
  Hubbard interaction. (b) The Chern number with SOC and 
  Hubbard interaction in the paramagnetic insulator phase.}
  \label{fig4}
\end{figure}

\textit{Quantum Anomalous Hall Effect with Paramagnetism.}---To calculate the spectrum function at any momentum in the first Brillouin zone, the twisted boundary conditions $\left\{\phi_x, \phi_y\right\}$ are employed~\cite{PhysRevB.44.9562}, which can be expressed as $\Psi\left(\mathbf{r}+N_\alpha\right)=\Psi\left(\mathbf{r}\right) 
\exp \left(i \phi_\alpha\right)$. The spectrum functions are defined at each 
grid point $\{i,j\}$ of the discretized phase space, where $\phi_\alpha = 2\pi 
i/N_\alpha$ (with $\alpha = x,y$) and $N_\alpha$ represents the number of 
subdivisions along each direction. in our calculation, we set $N_x=N_y=12$. By summing up
the spectrum functions, we obtain a refined density of states,  and find that the flat bands separated by a Mott-Hubbard gap of order $U$ around the Fermi level [see Fig.~\ref{fig4}(a)].

To quantify the topological characteristics of the paramagnetic ground state, we compute the Chern number by using a discretized form of the integration of Berry curvature~\cite{PhysRevB.31.3372, PhysRevB.103.035125, PhysRevB.104.195117, PhysRevLett.131.106601},
\begin{equation}
\mathcal{C}=\int \frac{d \phi_x d \phi_y}{2 \pi i}\left(\left\langle\partial_{\phi_x} 
\Psi^* | \partial_{\phi_y} \Psi\right\rangle-\left\langle\partial_{\phi_y} 
\Psi^* | \partial_{\phi_x} \Psi\right\rangle\right),
\end{equation}
where $\left\{\phi_x, \phi_y\right\}$ are employed in twisted boundary conditions. When $\lambda / t=0.02$ and $U/t\in\left[0.5, 1.1\right]$, there is a paramagnetic QAHE with $\mathcal{C}=2$, which gradually transits into a trivial paramagnetic Mott insulator phase for $U / t>1.1$. When $\lambda / t=0.1$ and $U/t\in\left[0.5, 1.1\right]$, the system exhibits a paramagnetic QAHE with $\mathcal{C}=6$, which further transits into a paramagnetic QAHE with $\mathcal{C}=2$ for $U/t\in\left[1.2, 2\right]$ and finally enters a trivial paramagnetic insulator phase for $U / t>2$. At $\lambda / t=0.2$, the initial paramagnetic QAHE with $\mathcal{C}=6$ ($U/t\in\left[0.5, 2.9\right]$) enters a paramagnetic QAHE with $\mathcal{C}=2$ ($U=3$) and finally becomes a trivial paramagnetic Mott insulator phase ($U/t\in\left[3.3, 6\right]$), respectively. It is noteworthy that when $U/t \in \left[3.1, 3.2\right]$, the band crossing occurs between the ground state and the first excited state bands. At $\lambda / t=0.3$, the initial paramagnetic QAHE with $\mathcal{C}=6$ ($U/t\in\left[0.5, 4\right]$) eventually transits into a trivial paramagnetic Mott insulator phase ($U/t\in\left[4.5, 6\right]$). When $U/t \in \left[4.1, 4.4\right]$, the ground state band and the first excited band maintain crossed. At $\lambda / t=0.4$, the system is a paramagnetic QAHE with $\mathcal{C}=6$ ($U/t\in\left[0.5, 4.9\right]$), and then the ground state band and the first excited band retain crossed when $U/t \in \left[5, 6\right]$ [see Fig.~\ref{fig4}(b)].

Taken together, we realize a paramagnetic QAHE in a Fermion-Hubbard model based on a dice lattice with weak SOC in the half-filled flat bands. The onsite Hubbard interaction stabilizes a ferrimagnetic phase, whereas the SOC drives the system into a paramagnetic QAHE phase, signifying 
concurrent magnetic and topological phase transitions. Additionally, to interpret the chirality of our system with spontaneous time-reversal symmetry breaking but without any magnetic order, the concept of loop current 
order in analogous kagome systems has been invoked~\cite{NM.20.1353,PhysRevB.104.045122,PhysRevB.104.035131,PhysRevB.104.075148,Nature.602.245,NP.18.1470,PhysRevLett.127.217601,PhysRevB.107.045127,PhysRevLett.131.086601,NC.14.678,NC.14.7845,JPhysSocJpn.93.033704,PhysRevLett.132.146501,Nature.631.60,ScienceBulletin.66.1384,PhysRevB.104.165136}.
If the loop current order is not considered, the mean filed approximation may lost topological properties of the 
system~\cite{PhysRevB.104.235115,j.isci.2023.107546}.

\textit{Summary and Discussion.}---We systematically investigate the magnetic and topological properties of a $2\times2$ unit-cell cluster of dice lattice with SOC and Hubbard interaction by using the exact diagonalization methods. At a fixed Hubbard interaction strength, along with the increase of SOC, the initial ferrimagnetic phase first transits into a paramagnetic phase, followed by 
ferro-, antiferro-, ferro-, and para-magnetic phases, respectively. Correspondingly, the ground state degeneracies for these phases are 5, 1, 2, 5, 2, and 1, respectively. We further demonstrate that the paramagnetic phase with weak SOC exhibits the QAHE, which shows nonuniform loop currents 
that break time-reversal symmetry and possesses a Chern number of $\mathcal{C}=2$ or $6$. With the half-filled flat bands, strong correlation effects produce a well-defined first excitation gap between the ground state and excited states, along with a clear Mott insulating gap near the Fermi level, which indicates the robustness of the paramagnetic QAHE against both thermal fluctuations and external perturbations.

Lastly, we provide a concise discussion on the experimental feasibility. The transition-metal oxide trilayer 
heterostructure $\text{SrTiO}_3/\text{SrIrO}_3/\text{SrTiO}_3$ exhibits both strong SOC on the $\text{Ir}_{4}^{+}$ ion and electronic correlation in half-filled flat bands~\cite{PhysRevB.84.241103}. 
Similarly, the twisted bilayers of 1T-$\text{ZrS}_2$ host ultra-strong SOC and flat 
bands~\cite{NC13.4915}. The hybrid topological phase has been reported in the $\alpha$-As crystal~\cite{Nature.628.527}. And other dice lattice systems provide additional platforms, such as $\text{VX}_2$ (X=CI, Br, I) monlayers~\cite{PhysRevLett.132.086802} and $\text{MoSi}_2\text{N}_4$ monolayer 
films~\cite{scienceabb7023}. By combining these recently discovered two-dimensional flat-band materials with SOC-driven quantum phase 
effects~\cite{CoshareScience.1.3,NP.15.443,PhysRevLett.124.183901,NC.11.4004,PhysRevB.102.125115,nanolett.2c00778,CP.5.198,
Nature.612.647,NP.19.1135,NRP.5.635,PhysRevB.110.L041121,CoshareScience.03.01}, the paramagnetic QAHE is highly expected to be observed.

\begin{acknowledgments}
We are grateful to Prof. Shun-Qing Shen for valuable discussions. This work was financially supported by National Key R$\&$D Program of China (2024YFA1408103), the National Natural Science Foundation of China (12474158, 12234017 and 12488101), Innovation Program for Quantum Science and Technology (2021ZD0302800), and Anhui Initiative in Quantum Information Technologies (AHY170000). We also thank the Supercomputing Center of University of Science and Technology of China for providing high-performance computing resources.
\end{acknowledgments}

$^\dagger$ These authors contribute equally to this work.


\begin{thebibliography}{999}	
\bibitem{PhysRevLett.45.494}{K. v. Klitzing, G. Dorda, and M. Pepper, Phys. Rev. Lett. {\bf 45}, 494 (1980).}
	
\bibitem{PhysRevB.23.5632}{R. B. Laughlin, Phys. Rev. B {\bf 23}, 5632(R) (1981).}
	
\bibitem{PhysRevLett.49.405}{D. J. Thouless, M. Kohmoto, M. P. Nightingale, and M. den Nijs, Phys. Rev. Lett. {\bf 49}, 405 (1982).}
	
\bibitem{PhysRevLett.61.2015}{F. D. M. Haldane, Phys. Rev. Lett. {\bf 61}, 2015 (1988).}

\bibitem{Science329}{R. Yu, W. Zhang, H. J. Zhang, S. C. Zhang, X. Dai, and Z. Fang, Science {\bf 329}, 61 (2010).}
	
\bibitem{PhysRevB.82.161414}{Z. H. Qiao, S. A. Yang, W. X. Feng, W. K. Tse, J. Ding, Y. G. Yao, J. Wang, and Q. Niu, Phys. Rev. B {\bf 82}, 161414(R) (2010).}
	
\bibitem{PhysRevB.87.085431}{W. D. Luo, and X. L. Qi, Phys. Rev. B {\bf 87}, 085431 (2013).}

\bibitem{CoshareScience.02.01}{S. Q. Shen, Coshare Science {\bf 02}, 01 (2024).} 

\bibitem{PhysRevLett.101.146802}{C. X. Liu, X. L. Qi, X. Dai, Z. Fang, and S. C. Zhang, Phys. Rev. Lett. {\bf 101}, 146802 (2008).}
	
\bibitem{science.340.167}{C. Z. Chang, J. S. Zhang, X. Feng, J. Shen, Z. C. Zhang, M. H. Guo, K. Li, Y. B. Ou, P. Wei, L. L. Wang, Z. Q. Ji, Y. Feng, S. H. Ji, X. Chen, J. F. Jia, X. Dai, Z. Fang, S. C. Zhang, K. He , Y. Y. Wang , L. Lu, X. C. Ma, and Q. K. Xue, Science {\bf 340}, 167 (2013).}
	
\bibitem{FOP.13.137308}{X. Z. Deng, H. L. Yang, S. F. Qi, X. H. Xu, and Z. H. 
Qiao, Front. Phys. {\bf 13}, 137308 (2018).}

\bibitem{science.365.605}{A. L. Sharpe, E. J. Fox, A. W. Barnard, J. Finney, K. Watanabe, T. Taniguchi, M. A. Kastner, and D. Goldhaber-Gordon, Science {\bf 365}, 605 (2019).}
	
\bibitem{science.367.900}{M. Serlin, C. L. Tschirhart, H. Polshyn, Y. Zhang, J. Zhu, K. Watanabe, T. Taniguchi, L. Balents, and A. F. Young, Science {\bf 367}, 900 (2019).}
	
\bibitem{Nature.588.610}{K. P. Nuckolls, M. Oh, D. Wong, B. Lian, K. Watanabe, T. Taniguchi, B. A. Bernevig, and A. Yazdani, Nature {\bf 588}, 610 (2020).}

\bibitem{PhysRevLett.126.117602} 
Y. F. Ren, et al., Phys. Rev. Lett. {\bf 126}, 117602 (2021).
	
\bibitem{PhysRevB.108.L161112}{G. Sethi, D. N. Sheng, and F. Liu, Phys. Rev. B {\bf 108}, L161112 (2023).}

\bibitem{Commun.Phys.6.240}{N. Mohanta, R. Soni, S. Okamoto, E. Dagotto, Commun Phys {\bf 6}, 240 (2023)}

\bibitem{PhysRevLett.129.036801}{Z. Li, Y. Han, and Z. Qiao, Phys. Rev. Lett. {\bf 129}, 036801 (2022).}

\bibitem{Nature.576.416}{M. M. Otrokov, I. I. Klimovskikh, H. Bentmann, D. Estyunin, A. Zeugner, Z. S. Aliev, S. Gaß, A. U. B. Wolter, A. V. Koroleva, A. M. Shikin, M. Blanco-Rey, M. Hoffmann, I. P. Rusinov, A. Yu. Vyazovskaya, S. V. Eremeev, Yu. M. Koroteev, V. M. Kuznetsov, F. Freyse, J. Sánchez-Barriga, I. R. Amiraslanov, M. B. Babanly, N. T. Mamedov, N. A. Abdullayev, V. N. Zverev, A. Alfonsov, V. Kataev, B. Büchner, E. F. Schwier, S. Kumar, A. Kimura, L. Petaccia, G. Di Santo, R. C. Vidal, S. Schatz, K. Kißner, M. Ünzelmann, C. H. Min, Simon Moser, T. R. F. Peixoto, F. Reinert, A. Ernst, P. M. Echenique, A. Isaeva, and E. V. Chulkov, Nature {\bf 576}, 416 (2019).}
	
\bibitem{PhysRevLett.122.107202}{M. M. Otrokov, I. P. Rusinov, M. Blanco-Rey, M. Hoffmann, A. Yu. Vyazovskaya, S. V. Eremeev, A. Ernst, P. M. Echenique, A. Arnau, and E. V. Chulkov, Phys. Rev. Lett. {\bf 122}, 107202 (2019).}

\bibitem{npjQuantum Mater.5.54}{I. I. Klimovskikh, M. M. Otrokov, D. Estyunin, S. V. Eremeev, S. O. Filnov, A. Koroleva, E. Shevchenko, V. Voroshnin, A. G. Rybkin, I. P. Rusinov, M. Blanco-Rey, M. Hoffmann, Z. S. Aliev, M. B. Babanly, I. R. Amiraslanov, N. A. Abdullayev, V. N. Zverev, A. Kimura, O. E. Tereshchenko, K. A. Kokh, L. Petaccia, G. D. Santo, A. Ernst, P. M. Echenique, N. T. Mamedov, A. M. Shikin, and E. V. Chulkov, npj Quantum Mater. {\bf 5}, 54 (2020).}
	
\bibitem{PhysRevB.100.121103}{J. H. Li, C. Wang, Z. T. Zhang, B. L. Gu, W. H. Duan, and Y. Xu, Phys. Rev. B {\bf 100}, 121103(R) (2019).}
	
\bibitem{SciAdv.5.eaaw5685}{J. H. Li, Y. Li, S. Q. Du, Z. Wang, B. L. Gu, S. C. Zhang, K. He, W. H. Duan, and Y. Xu, Sci. Adv. {\bf 5}, eaaw5685 (2019).}
	
\bibitem{PhysRevX.9.041038}{Y. J. Hao, P. F. Liu, Y. Feng, X. M. Ma, E. F. Schwier, M. Arita, S. Kumar, C. W. Hu, R. Lu, M. Zeng, Y. Wang, Z. Y. Hao, H. Y. Sun, K. Zhang, J. W. Mei, N. Ni, L. S. Wu, K. Shimada, C. Y. Chen, Q. H. Liu, and C. Liu, Phys. Rev. X {\bf 9}, 041038 (2019).}
	
\bibitem{PhysRevLett.122.206401}{D. Q. Zhang, M. J. Shi, T. S. Zhu, D. Y. Xing, H. J. Zhang, and J. Wang, Phys. Rev. Lett. {\bf 122}, 206401 (2019).}
	
\bibitem{PhysRevLett.123.096401}{H. Y. Sun, B. W. Xia, Z. J. Chen, Y. J. Zhang, P. F. Liu, Q. S. Yao, H. Tang, Y. J. Zhao, Hu Xu, and Q. H. Liu, Phys. Rev. Lett. {\bf 123}, 096401 (2019).}
	
\bibitem{PhysRevB.101.201408}{J. Li, J. Y. Ni, X. Y. Li, H. J. Koo, M. H. Whangbo, J. S. Feng, and H. J. Xiang, Phys. Rev. B {\bf 101}, 201408(R) (2020).}
	
\bibitem{NanoLett.20.2609}{P. M. Sass, W. B. Ge, J. Q. Yan, D. Obeysekera, J. J. Yang, and W. D. Wu, Nano Lett. {\bf 20}, 2609 (2020).}
	
\bibitem{PhysRevB.101.161109}{P. Swatek, Y. Wu, L. L. Wang, K. Lee, B. Schrunk, J. Q. Yan, and A. Kaminski, Phys. Rev. B {\bf 101}, 161109(R) (2020).}
		
\bibitem{PhysRevB.101.161113}{Y. Hu, L. X. Xu, M. Z. Shi, A. Luo, S. Peng, Z. Y. Wang, J. J. Ying, T. Wu, Z. K. Liu, C. F. Zhang, Y. L. Chen, G. Xu, X. H. Chen, and J. F. He, Phys. Rev. B {\bf 101}, 161113(R) (2020).}
	
\bibitem{PhysRevB.102.241406}{H. P. Sun, C. M. Wang, S. B. Zhang, R. Chen, Y. Zhao, C. Liu, Q. H. Liu, C. Y. Chen, H. Z. Lu, and X. C. Xie, Phys. Rev. B {\bf 102}, 241406(R) (2020).}
		
\bibitem{PhysRevLett.124.126402}{B. Lian, Z. C. Liu, Y. B. Zhang, and J. Wang, Phys. Rev. Lett. {\bf 124}, 126402 (2020).}
	
\bibitem{PhysRevB.103.245403}{Y. L. Han, S. Y. Sun, S. F. Qi, X. H. Xu, and Z. H. Qiao, Phys. Rev. B {\bf 103}, 245403 (2021).}

\bibitem{PhysRevLett.134.116603}{W. Liang, Z. Li, J. An, Y. Ren, Z. Qiao, and Q. Niu, Phys. Rev. Lett. {\bf 134}, 116603 (2025).}

\bibitem{PhysRevLett.100.156401}{S. Raghu, X. L. Qi, C. Honerkamp, and S. C. Zhang, Phys. Rev. Lett. {\bf 100}, 156401 (2008).}
	
\bibitem{PhysRevLett.117.096402}{W. Zhu, et al, Phys. Rev. Lett. {\bf 117}, 096402 (2016).}
	
\bibitem{PhysRevB.98.125144}{S. Sur, S. S. Gong, K. Yang, and O. Vafek, Phys. Rev. B {\bf 98}, 125144 (2018).}
	
\bibitem{PhysRevB.98.205146}{Y. F. Ren, T. S. Zeng, W. Zhu, and D. N. Sheng, Phys. Rev. B {\bf 98}, 205146 (2018).}
	
\bibitem{PhysRevB.106.205105}{H. Y. Lu, S. Sur, S. S. Gong, and D. N. Sheng, Phys. Rev. B {\bf 106}, 205105 (2022).}
	
\bibitem{PhysRevB.84.241103}{F. Wang and Y. Ran, Phys. Rev. B {\bf 84}, 241103(R) (2011).}

\bibitem{NC13.4915}{M. Claassen, L. Xian, D. M. Kennes, and A. Rubio, Nat Commun {\bf 13}, 4915 (2022).}

\bibitem{PhysRevLett.123.157601}{Y. D. Liao, Z. Y. Meng, and X. Y. Xu, Phys. Rev. Lett. {\bf 123}, 157601 (2019).}

\bibitem{ComputerPhysicsCommunications2017}{M. Kawamura, K. Yoshimi, T. Misawa, Y. Yamaji, S. Todo, and N. Kawashima, Computer Physics Communications {\bf 217}, 180 (2017).}
	
\bibitem{ComputerPhysicsCommunications2024}{K. Ido, M. Kawamura, Y. Motoyama, K. Yoshimi, Y. Yamaji, S. Todo, N. Kawashima, and T. Misawa, Computer Physics Communications {\bf 298}, 109093 (2024).}

\bibitem{PhysRevB.37.7359}{H. Q. Lin, J. E. Hirsch, and D. J. Scalapino, Phys. Rev. B {\bf 37}, 7359 (1988).}

\bibitem{PhysRevB.46.11779}{P. W. Leung, Z. P. Liu, E. Manousakis, M. A. Novotny, and P. E. Oppenheimer, Phys. Rev. B {\bf 46}, 11779 (1992).}

\bibitem{Eur.Phys.J.D.52.159}{F. López-Urías and G. M. Pastor, Eur. Phys. J. D {\bf 52}, 159-162 (2009).}

\bibitem{PhysRevX.5.041041}{J.P.F. LeBlanc, et al, Phys. Rev. X {\bf 5}, 041041 (2015).}
	
\bibitem{SupplementalMaterial}{See Supplemental Material at XXX for details of 
the dice lattice band structure in the non-interacting limit, density of states 
with Hubbard and SOC interactions, spin-dependent density with Hubbard and SOC 
interactions, a real-space plot of the spin magnetic moments, spin-spin 
correlations, a detailed analysis of Chern number in the paramagnetic QAHE, and 
density matrix renormalization group calculations. which includes Refs. 
[40,43-44,46,50-51,54-56,91]}

\bibitem{PhysRevLett.62.1201}{E. H. Lieb, Phys. Rev. Lett. {\bf 62}, 1201 (1989).}

\bibitem{PhysRevLett.72.1280}{S. Q. Shen, Z. M. Qiu, and G. S. Tian, Phys. Rev. Lett. {\bf 72}, 1280 (1994).}

\bibitem{PhysRevB.102.045105}{R. Soni, N. Kaushal, S. Okamoto, and E. Dagotto, Phys. Rev. B {\bf 102}, 045105 (2020).}

\bibitem{PhysRevB.44.9562}{D. Poilblanc, Phys. Rev. B {\bf 44}, 9562 (1991).}

\bibitem{PhysRevB.31.3372}{Q. Niu, D. J. Thouless, and Y. S. Wu, Phys. Rev. B {\bf 31}, 3372 (1985).}
				
\bibitem{PhysRevB.103.035125}{C. Shao, E. V. Castro, S. J. Hu, and R. Mondaini, Phys. Rev. B {\bf 103}, 035125 (2021).}
				
\bibitem{PhysRevB.104.195117}{T. C. Yi, S. J. Hu, E. V. Castro, and R. Mondaini, Phys. Rev. B {\bf 104}, 195117 (2021).}

\bibitem{PhysRevLett.131.106601}{J. C. Zhao, P. Z. Mai, B. Bradlyn, and P. Phillips, Phys. Rev. Lett. {\bf 131}, 106601 (2023).}



\bibitem{NM.20.1353}{Y. X Jiang, J. X. Yin, M. M. Denner, N. Shumiya, B. R. Ortiz, G. Xu, Z. Guguchia, J. Y. He, M. S. Hossain, X. X. Liu, J. Ruff, L. Kautzsch, S. S. Zhang, G. Q. Chang, I. Belopolski, Q. Zhang, T. A. Cochran, D. Multer, M. Litskevich, Z. J. Cheng, X. P. Yang, Z. Q. Wang, R. Thomale, T. Neupert, S. D. Wilson, and M. Z. Hasan, Nat. Mater. {\bf 20}, 1353–1357 (2021).}
	
\bibitem{PhysRevB.104.045122}{Y. P. Lin and R. M. Nandkishore, Phys. Rev. B {\bf 104}, 045122 (2021).}
	
\bibitem{PhysRevB.104.035131}{N. Shumiya, M. S. Hossain, J. X. Yin, Y. X. Jiang, B. R. Ortiz, H. X. Liu, Y. G. Shi, Q. W. Yin, H. C. Lei, S. S. Zhang, G. Q. Chang, Q. Zhang, T. A. Cochran, D. Multer, M. Litskevich, Z. J. Cheng, X. P. Yang, Z. Guguchia, S. D. Wilson, and M. Z. Hasan, Phys. Rev. B {\bf 104}, 035131 (2021).}
	
\bibitem{PhysRevB.104.075148}{Z. W. Wang, Y. X. Jiang, J. X. Yin, Y. K. Li, G. Y. Wang, H. L. Huang, S. Shao, J. J. Liu, P. Zhu, N. Shumiya, M. S. Hossain, H. X. Liu, Y. G. Shi, J. X. Duan, X. Li, G. Q. Chang, P. C. Dai, Z. J. Ye, G. Xu, Y. C. Wang, H. Zheng, J. F. Jia, M. Z. Hasan, and Y. G. Yao, Phys. Rev. B {\bf 104}, 075148 (2021).}
	
\bibitem{Nature.602.245}{C. Mielke III, D. Das, J. X. Yin, H. Liu, R. Gupta, Y. X. Jiang, M. Medarde, X. Wu, H. C. Lei, J. Chang, P. C. Dai, Q. Si, H. Miao, R. Thomale, T. Neupert, Y. Shi, R. Khasanov, M. Z. Hasan, H. Luetkens, and Z. Guguchia, Nature {\bf 602}, 245–250 (2022).}
	
\bibitem{NP.18.1470}{Y. S. Xu, Z. L. Ni, Y. Z. Liu, B. R. Ortiz, Q. W. Deng, S. D. Wilson, B. H. Yan, L. Balents, and L. Wu, Nat. Phys. {\bf 18}, 1470–1475 (2022).}
	
\bibitem{PhysRevLett.127.217601}{M. M. Denner, R. Thomale, and T. Neupert, Phys. Rev. Lett. {\bf 127}, 217601 (2022).}
	
\bibitem{PhysRevB.107.045127}{J. W. Dong, Z. Q. Wang, and S. Zhou, Phys. Rev. B {\bf 107}, 045127 (2023).}
	
\bibitem{PhysRevLett.131.086601}{J. J. Zeng, Q. M. Li, X. Yang, D. H. Xu, and R. Wang, Phys. Rev. Lett. {\bf 131}, 086601 (2023).}
	
\bibitem{NC.14.678}{G. L. Zheng, C. Tan, Z. Chen, M. Wang, X. D. Zhu, S. Albarakati, M. Algarni, J. Partridge, L. Farrar, J. H. Zhou, W. Ning, M. L. Tian, M. S. Fuhrer, and L. Wang, Nat Commun {\bf 14}, 678 (2023).}
	
\bibitem{NC.14.7845}{R. Tazai, Y. Yamakawa, and H. Kontani, Nat Commun {\bf 14}, 7845 (2023).}
	
\bibitem{JPhysSocJpn.93.033704}{K. Shimura, R. Tazai, Y. Yamakawa, S. Onari, and H. Kontani, J. Phys. Soc. Jpn. {\bf 93}, 033704 (2024).}
	
\bibitem{PhysRevLett.132.146501}{H. Q. Li, Y. B. Kim, and H. Y. Kee, Phys. Rev. Lett. {\bf 132}, 146501 (2024)}
	
\bibitem{Nature.631.60}{Y. Q. Xing, S. Bae, E. Ritz, F. Yang, T. Birol, A. N. C. Salinas, B. R. Ortiz, S. D. Wilson, Z. Q. Wang, R. M. Fernandes, and V. Madhavan, Nature {\bf 631}, 60–66 (2024).}

\bibitem{ScienceBulletin.66.1384}{X. L. Feng, K. Jiang, Z. Q. Wang, and J. P. Hu, Science Bulletin {\bf 66}, 1384 (2021).}

\bibitem{PhysRevB.104.165136}{X. L. Feng, Y. Zhang, K. Jiang, and J. P. Hu, Phys. Rev. B {\bf 104}, 165136 (2021).}

\bibitem{PhysRevB.104.235115}{R. Soni, A. B. Sanyal, N. Kaushal, S. Okamoto, A. Moreo, and E. Dagotto, Phys. Rev. B {\bf 104}, 235115 (2021).}

\bibitem{j.isci.2023.107546}{S. Q. Lin, H. Tan, P. H. Fu, and J. F. Liu, iScience {\bf 26}, 107546 (2023).}
	
\bibitem{Nature.628.527}{M. S. Hossain, F. Schindler, R. Islam, Z. Muhammad, Y. X. Jiang, Z. J. Cheng, Q. Zhang, T. Hou, H. Chen, M. Litskevich, B. Casas, J. X. Yin, T. A. Cochran, M. Yahyavi, X. P. Yang, L. Balicas, G. Q. Chang, W. S. Zhao, T. Neupert, and M. Z. Hasan, Nature {\bf 628}, 527–533 (2024).}	

\bibitem{PhysRevLett.132.086802}{C. Liu, W. Ren, and S. Picozzi, Phys. Rev. Lett. {\bf 132}, 086802 (2024).}

\bibitem{scienceabb7023}{Y. L. Hong, Z. B. Liu, L. Wang, T. Y. Zhou, W. Ma, C. Xu, S. Feng, L. Chen, M. L. Chen, D. M. Sun, X. Q. Chen, H. M. Cheng, and W. C. Ren, Science {\bf 369}, 670 (2020).}

\bibitem{CoshareScience.1.3}{F. Liu, Coshare Science {\bf 01}, v3, 1 (2023).}

\bibitem{NP.15.443}{J. X. Yin, S. S. Zhang, G. Q. Chang, Q. Wang, S. S. Tsirkin, Z. Guguchia, B. Lian, H. B. Zhou, K. Jiang, I. Belopolski, N. Shumiya, D. Multer, M. Litskevich, T. A. Cochran, H. Lin, Z. Q. Wang, T. Neupert, S. Jia, H. C. Lei, and M. Z. Hasan, Nature Physics {\bf 15}, 443 (2019).}

\bibitem{PhysRevLett.124.183901}{J. Ma, J. W. Rhim, L. Q. Tang, S. Q. Xia, H. P. Wang, X. Y. Zheng, S. Q. Xia, D. H. Song, Y. Hu, Y. G. Li, B. J. Yang, D. Leykam, and Z. G. Chen, Phys. Rev. Lett. {\bf 124}, 183901 (2020).}

\bibitem{NC.11.4004}{M. G. Kang, S. Fang, L. D. Ye, H. C. Po, J. Denlinger, C. Jozwiak, A. Bostwick, E. Rotenberg, E. Kaxiras, J. G. Checkelsky, R. Comin, Nat Commun {\bf 11}, 4004 (2020).}

\bibitem{PhysRevB.102.125115}{Y. Zhou, G. Sethi, C. Zhang, X. J. Ni, and F. Liu, Phys. Rev. B {\bf 102}, 125115 (2020).}

\bibitem{nanolett.2c00778}{Z. Y. Sun, H. Zhou, C. X. Wang, S. Kumar, D. Y. Geng, S. S. Yue, X. Han, Y. Haraguchi, K. Shimada, P. Cheng, L. Chen, Y. Shi, K. Wu, S. Meng, and B. Feng, Nano Letters {\bf 22} 4596 (2022).}

\bibitem{CP.5.198}{S. Okamoto, N. Mohanta, E. Dagotto, and D. N. Sheng, Commun Phys {\bf 5}, 198 (2022).}

\bibitem{Nature.612.647}{J. X. Yin, B. Lian, and M. Z. Hasan, Nature {\bf 612}, 647 (2022).}
	
\bibitem{NP.19.1135}{D. D. Sante, C. Bigi, P. Eck, S. Enzner, A. Consiglio, G. Pokharel, P. Carrara, P. Orgiani, V. Polewczyk, J. Fujii, P. D. C. King, I. Vobornik, G. Rossi, I. Zeljkovic, S. D. Wilson, R. Thomale, G. Sangiovanni, G. Panaccione, F. Mazzola, Nature Physics {\bf 19}, 1135 (2023).}

\bibitem{NRP.5.635}{Y. Wang, H. Wu, G. T. McCandless, J. Y. Chan, and M. N. Ali, Nat Rev Phys {\bf 5}, 635–658 (2023).}

\bibitem{PhysRevB.110.L041121}{Y. P. Lin, C. X. Liu, and J. E. Moore, Phys. 
Rev. B {\bf 110}, L041121 (2024).}

\bibitem{CoshareScience.03.01}{S. Hasegawa, Coshare Science {\bf 03}, 01 
(2025).}

\bibitem{SciPost Phys. Codebases 4}{M. Fishman, S. R. White, and E. M. 
Stoudenmire, SciPost Phys. Codebases 4 (2022) .}
\end{thebibliography}
\end{document}